# Support sufficiency as action-sufficient compression: a single-cycle rate-regret formulation

Mark Walsh, PharmD
Corresponding email: mark.walsh@rutgers.edu

**Abstract**

Robust decision-making requires compression. A system that forms a rich support state cannot usually preserve its full structure at the point of action. It must retain only those distinctions needed to act, verify, abstain, or defer under the current consequence geometry. This paper formalizes support sufficiency as action-sufficient compression. Let $H$ denote a full support state, $\mathcal{A}$ a finite action set, and $Z$ a consequence geometry specifying payoff structure. For fixed $Z$, the coarsest exactly action-sufficient compression is the quotient of support space by policy equivalence. Two support states may be merged exactly when they require the same optimal action. This clarifies why content-only and scalar-confidence-only arbitration fail whenever their induced partitions cross action boundaries. Approximate sufficiency is then defined by bounded expected policy regret. In the finite single-cycle setting, this yields a rate-regret problem with source $H$, reproduction alphabet $\mathcal{A}$, and distortion given by consequence-sensitive regret. The optimal stochastic action channel inherits the standard rate-distortion Gibbs form, applied here to support states with regret distortion. The contribution is interpretive: action adequacy is distinguished from reconstruction fidelity, information-bottleneck prediction, and rational inattention. Robust single-cycle arbitration does not require preserving all support, but it does require preserving the distinctions that consequence geometry makes action-relevant.

**Keywords:** Support sufficiency, action-sufficient compression, rate-distortion theory, policy regret, information bottleneck, bounded rationality, belief arbitration

## 1. Introduction

Decision-making systems often form richer internal states than they can preserve at the point of action. A perceptual, cognitive, or artificial system may not only select a content state, but also carry information about how that content was supported, including conflict among sources, provenance, reliability, drift, or verification value. Yet when the system must act, verify, abstain, or defer, much of this support structure must be compressed. The central question is therefore not whether compression occurs, but which distinctions must survive compression for action to remain adequate.

This question is especially natural in cognitive science, where confidence, uncertainty, conflict, and reliability signals are increasingly treated as control-relevant variables rather than passive annotations on content (Fleming and Daw 2017; Pouget et al. 2016). Perceptual and metacognitive systems do not merely estimate what is present. They also regulate whether to commit, sample again, withhold action, or revise a belief. A scalar confidence estimate can support some of this control, but it may collapse distinct uncertainty structures that call for different policies (Pouget et al. 2016). Low precision, cross-source conflict, provenance fragility, and drift can all produce similar apparent confidence while requiring different patterns of action, verification, or abstention. The present formalism isolates this issue at the arbitration bottleneck by asking what compressed information must remain available for action to stay adequate once a rich support state has been formed.

Classical rate-distortion theory studies how much information about a source must be preserved to satisfy a fidelity criterion (Shannon 1959; Berger 1971). In the present setting, the fidelity criterion is not reconstruction of the original support state. It is action adequacy. A compression is successful when it preserves the action that should be taken, or when the expected regret induced by compression remains within tolerance. This reframes support preservation as a consequence-sensitive compression problem rather than a demand to retain all upstream detail.

The formal structure developed below is adjacent to several established literatures. In decision theory, the exact partition result is closely related to the general idea that a representation is sufficient for a decision problem when it preserves the decision-relevant information induced by the payoff structure. This lineage includes Blackwell's comparison of experiments and later work on the value of information (Blackwell 1953). In economics and bounded-rationality theory, rational inattention and information-theoretic decision models likewise study tradeoffs between utility and information cost (Matějka and McKay 2015; Tishby and Polani 2011; Ortega and Braun 2013; Genewein et al. 2015; Caplin and Dean 2015). The present contribution is therefore not the discovery of a new generic information-theoretic solution. It is the identification of support-sufficient arbitration as a particular decision problem in which the source is evidential support structure and the distortion is consequence-sensitive policy regret.

This paper builds on earlier work arguing that selected content and scalar confidence can be insufficient for robust arbitration. That work introduced the idea that globally available support structure can accompany content and shape confidence, report, and control (Walsh 2026d); showed computationally that support-conditioned calibration can improve behavior under regime shift (Walsh 2026a); generalized support into the geometry of evidence, conflict, provenance, and drift (Walsh 2026b); and developed a recurrent model in which support resolution is regulated by consequence geometry, memory, and resource cost (Walsh 2026e). The present paper extracts the single-cycle formal core behind those claims.

Let $H$ denote a full support state, $\mathcal{A}$ a finite action set, and $Z$ a consequence geometry specifying the payoff structure of the current decision problem. A compression $g$ is action-sufficient when the policy selected from $g(H)$ matches, or approximates within tolerance, the policy that would have been selected from the full support state.

The first result is exact. For fixed $Z$, the coarsest exactly action-sufficient compression is the quotient of support space by policy equivalence. In plain terms, two support states may be placed in the same compression cell if and only if they require the same optimal action. The second result is approximate. When bounded regret is allowed, support sufficiency becomes a rate-regret problem. The system minimizes information retained about $H$ subject to an upper bound on expected policy regret. Because the action set is finite, the approximate single-cycle problem reduces to a rate-distortion problem over noisy action recommendations. The relevant distortion is not reconstruction error, but policy loss.

The contribution is therefore interpretive and formal rather than a new generic decision-theoretic theorem. First, the paper identifies the decision-sufficiency quotient as the exact object corresponding to support sufficiency in a single-cycle arbitration problem. Second, it shows that content-only and scalar-confidence-only compression fail when their cells contain support-sensitive action boundaries. Third, it specializes rate-distortion machinery to this setting by treating full support $H$ as the source, actions as the reproduction alphabet, and consequence-sensitive policy regret as the distortion.

## 2. Formal setup

### Definition 1. Support state, action set, and consequence geometry

Let $H \in \mathcal{H}$ denote the full support state available prior to arbitration. $H$ may include integrated evidence strength, conflict structure, provenance fragility, drift indicators, memory-conditioned context, or any other structure relevant to policy.

Let $\mathcal{A}$ denote a finite action set. In the simplest case, $\mathcal{A} = \{\text{act,verify,abstain}\}$.

Let $Z$ denote the consequence geometry. $Z$ specifies the action-relevant payoff structure, including error asymmetries, verification value, abstention cost, and any other features determining which action is rational under a given support state.

Unless otherwise stated, the main single-cycle results below are stated for finite support spaces and finite action sets. The exact quotient construction in Proposition 1 requires only a well-defined optimal policy and therefore extends more generally. The deterministic partition and rate-regret results are stated in the finite setting, where cell sums and the standard finite rate-distortion specialization apply.

### Definition 2. Full-information utility and optimal policy

For each action $a \in \mathcal{A}$, support state $H \in \mathcal{H}$, and consequence geometry $Z$, let $U(a \mid H, Z)$ denote the expected utility of taking action $a$ given $H$ and $Z$.

The full-information optimal policy is

$$\pi^*(H, Z) = \arg\max_{a \in \mathcal{A}} U(a \mid H, Z). \tag{1}$$

Unless otherwise stated, ties are resolved by a fixed tie-breaking rule. Equivalently, $\pi^*$ may be treated as set-valued, but fixed tie-breaking keeps the first-pass formulation simple.

### Definition 3. Compression map and compression cells

A support compression is a map $g: \mathcal{H} \to \mathcal{S}$ where $S = g(H)$ is the compressed support code available to arbitration. The compression $g$ induces a partition of $\mathcal{H}$ into preimage cells,

$$g^{-1}(s) = \{H \in \mathcal{H}: g(H) = s\}. \tag{2}$$

A preimage cell is the set of all full support states that are assigned the same compressed code. Once states are placed in the same cell, the compressed policy can no longer distinguish them.

Let $P(H)$ denote a probability distribution over $\mathcal{H}$. This distribution is used to define conditional expected utilities under compression and expected regret under approximate sufficiency.

**Definition 4. Compressed utility and compressed policy**

Because a compressed state $S = s$ may correspond to many full support states, for any compressed state $s$ with $P(g(H) = s) > 0$, define the compressed utility as

$$U_g(a \mid s, Z) = \mathbb{E}[U(a \mid H, Z) \mid g(H) = s]. \tag{3}$$

The compressed policy induced by $g$ is

$$\pi_g(s, Z) = \arg\max_{a \in \mathcal{A}} \; U_g(a \mid s, Z). \tag{4}$$

For a particular full support state $H$, the action selected under compression is $\pi_g(g(H), Z)$.

Equations (3) and (4) define the expected-utility compressed policy on positive-probability cells. Null cells are irrelevant for expected-regret calculations. When pointwise exact sufficiency is evaluated over all $H \in \mathcal{H}$, the compressed policy is understood as defined directly on all compression cells, with the required action specified wherever exact equality is being tested.

**Definition 5. Exact action sufficiency**

A compression $g: \mathcal{H} \to \mathcal{S}$ is exactly action-sufficient for consequence geometry $Z$ if

$$\pi_g(g(H), Z) = \pi^*(H, Z) \quad \text{for all } H \in \mathcal{H}. \tag{5}$$

Exact action sufficiency means that compression never changes the selected action relative to the full-information policy.

**Definition 6. Policy equivalence**

For fixed $Z$, define the relation $\sim_Z$ on $\mathcal{H}$ by

$$H \sim_Z H' \Leftrightarrow \pi^*(H, Z) = \pi^*(H', Z). \tag{6}$$

The equivalence classes of $\sim_Z$ partition support space into regions whose members require the same optimal action under $Z$.

Together, equations (1) through (6) define the single-cycle decision object: full-information policy, compression, compressed policy, exact action sufficiency, and policy equivalence.

## 3. The coarsest exact action-sufficient partition

Equation (6) defines the policy-equivalence relation induced by a fixed consequence geometry $Z$. The next result states that the quotient over those equivalence classes is the exact compression object for single-cycle action sufficiency.

**Proposition 1. Decision-sufficiency quotient for support states**

For fixed consequence geometry $Z$, the coarsest exactly action-sufficient compression is the quotient map

$$q_Z : \mathcal{H} \to \mathcal{H}/\sim_Z, \tag{7}$$

which maps each support state $H$ to its policy-equivalence class $[H]_Z$.

**Proof**

Suppose $H, H' \in \mathcal{H}$ satisfy $\pi^*(H, Z) \neq \pi^*(H', Z)$.

Any exactly action-sufficient compression must separate $H$ and $H'$. If it did not, then $g(H) = g(H')$. The compressed policy would receive the same compressed input for both states and would therefore select the same action for both. It could not equal two different full-information optimal actions. Exact action sufficiency would fail.

Therefore, any exactly action-sufficient compression must distinguish all pairs of support states that lie in different $\sim_Z$ equivalence classes.

Conversely, the quotient map $q_Z$ collapses only those support states that share the same full-information optimal action. For every $H$, all states in $[H]_Z$ have the same optimal action by construction. A policy defined on the quotient classes can therefore select that action and satisfy $\pi_{q_Z}(q_Z(H), Z) = \pi^*(H, Z)$ for all $H \in \mathcal{H}$.

Thus $q_Z$ is exactly action-sufficient, and no coarser compression can be exactly action-sufficient. ■

**Interpretation**

The fundamental object of support compression is not the full belief state, confidence, or support in general. It is the partition of support space into action-equivalent regions. Support must survive compression only to the extent needed to keep the system on the correct side of action boundaries. This formalizes the support-sufficiency claim that the goal is not maximal retention of support, but preservation of policy-relevant differentiation. This quotient construction is not meant to be a new generic theorem about decision problems. It is the support-sufficiency version of a familiar decision-theoretic idea. Once utilities and actions are fixed, exact compression must preserve the distinctions that affect the optimal decision. Its role here is to identify that object for support states containing evidential strength, conflict, provenance, drift, and memory-conditioned context.

## 4. Constructive example: equal confidence, different action

Proposition 1 is structural. A short example shows why the quotient partition can be nontrivially finer than scalar confidence.

Let two support states have the same selected content and the same scalar confidence, $x(H_1) = x(H_2)$ and $c(H_1) = c(H_2) = 0.90$.

Let $H_1$ be a high-confidence state with clean provenance, and let $H_2$ be a high-confidence state with fragile provenance. Suppose the system is operating in a threat regime $Z_{\text{threat}}$, where mistaken action is costly and verification has positive value. Table 1 gives a minimal utility specification in which these equal-confidence states require different optimal actions.

| **Support state** | **Scalar confidence** | **Provenance** | $U(\text{act} \mid H, Z)$ | $U(\text{verify} \mid H, Z)$ | $U(\text{abstain} \mid H, Z)$ | **Optimal action** |
|---|---|---|---|---|---|---|
| $H_1$ | 0.90 | Clean | 0.50 | 0.35 | 0.00 | Act |
| $H_2$ | 0.90 | Fragile | -0.10 | 0.35 | 0.00 | Verify |

**Table 1. Equal confidence with different optimal actions under threat-sensitive consequences.** Both support states have the same selected content and scalar confidence, $c = 0.90$, but differ in provenance. Under $Z_{\text{threat}}$, clean provenance supports acting, while fragile provenance makes verification the optimal action. The example illustrates how confidence-only compression can merge support states that lie in different policy-equivalence classes.

As Table 1 shows, both states have the same content and confidence. A confidence-only compression therefore places them in the same cell: $g_c(H_1) = g_c(H_2) = 0.90$. But they lie in different policy-equivalence classes, $\pi^*(H_1, Z_{\text{threat}}) = \text{act}$ and $\pi^*(H_2, Z_{\text{threat}}) = \text{verify}$.

Thus confidence-only compression crosses an action boundary. If the compressed policy chooses act, it incurs regret on $H_2$. If it chooses verify, it incurs regret on $H_1$. A support-aware compression that preserves provenance status separates the two states and achieves exact action sufficiency in this example.

This example is not meant to prove that every confidence signal fails. It demonstrates the substantive condition under which confidence fails. Equal-confidence states can require different actions when consequence geometry is sensitive to support structure beyond scalar confidence.

## 5. A natural condition for scalar-confidence failure

The constructive example supplies a witness. More generally, scalar confidence fails when a confidence cell contains support states that lie on opposite sides of an action boundary. In that case, confidence-only compression does not merely omit descriptive detail. It induces policy regret by forcing one action onto states that require different actions.

**Proposition 2. Support-sensitive action boundaries induce confidence-compression regret**

Let $c: \mathcal{H} \to \mathbb{R}$ be a scalar confidence map, and let $g_c(H) = c(H)$ be confidence-only compression. Suppose that for some confidence value $c_0$, there are support states $H_1 = H(c_0, r_1)$ and $H_2 = H(c_0, r_2)$ with positive probability such that $c(H_1) = c(H_2) = c_0$ but $\pi^*(H_1, Z) \neq \pi^*(H_2, Z)$.

Let $C_{c_0} = \{H \in \mathcal{H} : c(H) = c_0\}$ be the confidence cell containing these states. Then confidence-only compression incurs cell regret

$$\Delta(C_{c_0}; Z) = \sum_{H \in C_{c_0}} P(H) \max_{a \in \mathcal{A}} U(a \mid H, Z) - \max_{a \in \mathcal{A}} \sum_{H \in C_{c_0}} P(H) U(a \mid H, Z). \tag{8}$$

Equation (8) is the value lost when the full-information policy may vary within a confidence cell but the compressed policy must choose a single action for the whole cell. If $\Delta(C_{c_0}; Z) > 0$, then scalar confidence is not exactly action-sufficient for $Z$. If the total regret induced by confidence cells exceeds $\varepsilon$, then scalar confidence is not $\varepsilon$-action-sufficient for $Z$. Different full-information actions within a confidence cell are sufficient for failure of pointwise exact action sufficiency under the fixed tie-breaking convention. Positive cell regret is the stronger utility-loss condition needed for regret-based failure.

**Proof**

Because $g_c(H) = c(H)$, all states in $C_{c_0}$ are mapped to the same compressed code. The compressed policy must therefore choose one action for the entire confidence cell. The full-information policy, by contrast, can choose different actions for different support states inside the cell. The difference between the full-information value and the best single-action value for the cell is exactly $\Delta(C_{c_0}; Z)$. If this quantity is positive, confidence-only compression loses value by merging support states that lie on different sides of an action boundary. Exact sufficiency fails. If the sum of such cell regrets exceeds $\varepsilon$, approximate sufficiency also fails. ■

**Interpretation**

This proposition is the support-specific hinge of the paper. It identifies when the support structure inside $H$ matters formally rather than only interpretively. Conflict, provenance fragility, drift, or other support dimensions matter when they create action boundaries inside a scalar-confidence cell. In that case, confidence-only compression does not merely omit descriptive detail. It induces positive policy regret. Achieving a lower regret tolerance then requires preserving information beyond scalar confidence, either by refining the confidence cell or by carrying a support code that tracks the relevant dimension.

A simple sufficient condition for positive cell regret is a sign change in a utility gap along a support dimension not encoded by confidence. For example, if

$$D_Z(r) = U(\, a \mid H(c_0, r), Z\,) - U(\, b \mid H(c_0, r), Z\,) \tag{9}$$

satisfies $D_Z(r_1) > 0$ and $D_Z(r_2) < 0$ then varying $r$ at fixed confidence $c_0$ moves the support state across an $a$-versus-$b$ action boundary. When those states have positive probability and the resulting confidence cell has positive regret, scalar confidence fails as an action-sufficient compression.

## 6. Content-only and scalar-confidence insufficiency

Let $x: \mathcal{H} \to \mathcal{X}$ be a content-selection map. A content-only compression has the form $g_x(H) = x(H)$.

### Corollary 1. Content-only insufficiency

Suppose there exist $H_1, H_2 \in \mathcal{H}$ and consequence geometry $Z$ such that $x(H_1) = x(H_2)$ but $\pi^*(H_1, Z) \neq \pi^*(H_2, Z)$.

Then content-only compression is not exactly action-sufficient for $Z$.

### Proof

Because $x(H_1) = x(H_2)$, content-only compression maps $H_1$ and $H_2$ to the same compressed state. The compressed policy must therefore select the same action for both. But the full-information optimal actions differ. Therefore content-only compression merges states across an action boundary and fails exact action sufficiency. ■

Scalar-confidence-only compression is the special case addressed more strongly by Proposition 2. Let $c: \mathcal{H} \to \mathbb{R}$ be a scalar confidence map, and let $g_c(H) = c(H)$.

If a confidence cell contains support states that require different optimal actions, then confidence-only compression merges states across an action boundary. If the induced cell regret $\Delta\left(C_{c_0}; Z\right)$ is positive, exact action sufficiency fails. In that case, if the total regret across confidence cells exceeds $\varepsilon$, then scalar-confidence-only compression is not $\varepsilon$-action-sufficient.

### Interpretation

This does not show that scalar confidence is useless. It shows that confidence is sufficient only when its level sets refine, or approximate closely enough, the action-equivalence partition induced by $Z$. Scalar confidence can be adequate in benign regimes while failing in threat, opportunity, or drift-sensitive regimes where collapsed support distinctions carry higher consequence.

## 7. Deterministic approximate sufficiency

Exact action preservation is useful as a boundary case, but bounded systems usually require acceptable control rather than perfect action preservation.

**Definition 7. Compression regret**

The regret of using compression $g$ at full support state $H$ under consequence geometry $Z$ is

$$R_g(H, Z) = U(\pi^*(H, Z) \mid H, Z) - U\big(\pi_g(g(H), Z) \mid H, Z\big). \tag{10}$$

**Definition 8. $\varepsilon$-action sufficiency**

For a distribution $P(H)$, a compression $g$ is $\varepsilon$-action-sufficient for $Z$ if

$$\mathbb{E}_{\mathrm{H}}\big[R_g(H, Z)\big] \leq \varepsilon. \tag{11}$$

Equations (10) and (11) shift the criterion from exact action preservation to bounded expected regret. The question is no longer whether compression preserves every action boundary exactly, but whether the expected control loss induced by compression remains within tolerance.

**Proposition 3. Finite-state bounded-regret partition characterization**

Let $\mathcal{H}$ and $\mathcal{A}$ be finite. Any deterministic compression $g: \mathcal{H} \to \mathcal{S}$ induces a partition $\mathcal{P}_g$ of support space. For each cell $C \in \mathcal{P}_g$, define the cell regret

$$\Delta(C; Z) = \sum_{H \in C} P(H) \max_{a \in \mathcal{A}} U\,(a \mid H, Z) - \max_{a \in \mathcal{A}} \sum_{H \in C} P(H) U\,(a \mid H, Z). \tag{12}$$

Then $g$ is $\varepsilon$-action-sufficient for $Z$ if and only if

$$\sum_{C \in \mathcal{P}_g} \Delta(C; Z) \leq \varepsilon. \tag{13}$$

For any chosen partition-complexity functional $\mathcal{C}$, a minimal deterministic bounded-regret compression is any partition solving

$$\min_{\mathcal{P}} \mathcal{C}(\mathcal{P}) \quad \text{subject to} \quad \sum_{C \in \mathcal{P}} \Delta(C; Z) \leq \varepsilon, \tag{14}$$

where $\mathcal{C}(\mathcal{P})$ measures partition complexity. Thus equation (13) gives the bounded-regret feasibility condition, while equation (14) states the corresponding deterministic compression problem.

**Proof**

Within a cell $C$, the compressed policy must choose one action for every $H \in C$. The best such action is

$$a_C^* = arg \max_{a \in \mathcal{A}} \sum_{H \in C} P(H) U(a \mid H, Z).$$

The full-information value contributed by the states in $C$ is

$$\sum_{H \in C} P(H) \max_{a \in \mathcal{A}} U(a \mid H, Z).$$

The best compressed value contributed by $C$ is

$$\max_{a \in \mathcal{A}} \sum_{H \in C} P(H) U(a \mid H, Z).$$

The difference is exactly $\Delta(C; Z)$. Summing over all cells gives the total expected regret of the partition. Therefore the compression is $\varepsilon$-action-sufficient if and only if the sum of its cell regrets is at most $\varepsilon$. ■

**Note on approximate equivalence**

One should not define approximate support classes only by pairwise mergeability. It may be acceptable to merge $H_1$ with $H_2$, and acceptable to merge $H_2$ with $H_3$, while merging all three states produces regret above tolerance. Pairwise approximate equivalence need not be transitive. The partition-level criterion in equation (13) avoids this problem by evaluating regret over whole compression cells rather than only pairwise merges.

## 8. Action-channel reduction

The deterministic partition proposition solves the finite deterministic case. The stochastic case gives a stronger information-theoretic form.

Suppose a compressed support code $S$ is ultimately decoded, possibly stochastically, into an action recommendation $\hat{A}$, understood as the action variable used for one-step regret evaluation. Then the single-cycle decision process has the Markov structure

$$H \to S \to \hat{A}. \tag{15}$$

By data processing (Cover and Thomas 2006),

$$I(H; \hat{A}) \leq I(H; S). \tag{16}$$

The Markov structure in equation (15), together with the data-processing inequality in equation (16), shows that the induced action recommendation cannot carry more information about $H$ than the support code that produced it. Moreover, in the single-cycle problem, regret depends only on the action actually selected, not on additional information in $S$ that does not affect action.

Therefore, for the purposes of one-step expected regret and information rate, any support code $S$ can be replaced by its induced action recommendation $\hat{A}$ without increasing information and without changing immediate policy regret. This reduces approximate support compression to the induced noisy action channel,

$$p(\hat{a} \mid H). \tag{17}$$

Equation (17) is the action-channel form of the single-cycle problem. At this level, the system does not need to reconstruct $H$. It only needs to preserve enough information about $H$ to choose an action with acceptable regret.

## 9. Rate-regret formulation of approximate sufficiency

### Definition 9. Regret distortion

Define the regret distortion of selecting action $a$ at support state $H$ under consequence geometry $Z$ as

$$d_Z(H, a) = U(\pi^*(H, Z) \mid H, Z) - U(a \mid H, Z). \tag{18}$$

This distortion is zero when $a$ is optimal and positive when $a$ is suboptimal.

### Definition 10. Rate-regret function

For finite $\mathcal{H}$ and finite $\mathcal{A}$, define the rate-regret function

$$R_Z(\varepsilon) = \min_{p(\hat{a} \mid H)} I\left(H; \hat{A}\right) \quad \text{subject to} \quad \mathbb{E}\left[d_Z\left(H, \hat{A}\right)\right] \leq \varepsilon. \tag{19}$$

Here $H$ is the source variable, $\hat{A}$ is the action recommendation, and $d_Z$ is policy regret. Equation (19) is the rate-regret analogue of the classical rate-distortion problem. The source is $H$, the reproduction variable is the action recommendation $\hat{A}$, and the distortion is the regret distortion defined in equation (18).

### Proposition 4. Standard rate-distortion specialization under regret distortion

For finite $\mathcal{H}$ and finite $\mathcal{A}$, the Lagrange-optimal stochastic action channel for the rate-regret problem has the form

$$p_\beta(\hat{a} \mid H) = \frac{p_\beta(\hat{a}) \exp[-\beta d_Z(H, \hat{a})]}{\sum_{a' \in \mathcal{A}} p_\beta(a') \exp[-\beta d_Z(H, a')]}, \tag{20}$$

where $\beta \geq 0$ is the Lagrange multiplier controlling the regret-information tradeoff, and

$$p_\beta(\hat{a}) = \sum_H P(H) p_\beta(\hat{a} \mid H) \tag{21}$$

is determined self-consistently.

**Derivation and standard form**

This follows from the standard variational derivation of the rate-distortion function for a finite source and finite reproduction alphabet (Shannon 1959; Berger 1971; Cover and Thomas 2006). The constrained problem in equation (19) is a rate-distortion problem with source $H$, reproduction alphabet $\mathcal{A}$, and distortion $d_Z$. Introducing the Lagrange multiplier $\beta$ gives the unconstrained objective

$$\mathcal{L} = I(H; \hat{A}) + \beta \mathbb{E}[d_Z(H, \hat{A})]. \tag{22}$$

Minimizing equation (22), with normalization imposed for each conditional distribution $p(\hat{a} \mid H)$, yields the Gibbs channel in equation (20). The marginal $p_\beta(\hat{a})$ must satisfy the self-consistency condition in equation (21). Sweeping $\beta$ traces the rate-regret frontier.

The present contribution is not the Gibbs form itself, which is standard in rate-distortion theory. The contribution is the support-sufficiency specialization, where the source is the full support state, the reproduction alphabet is the action set, and the distortion measure is consequence-sensitive policy regret.

**Remark. Computability by Blahut-Arimoto iteration**

For fixed $\beta$, the self-consistency condition in equation (21) can be solved by the standard Blahut-Arimoto iteration for rate-distortion problems (Blahut 1972; Arimoto 1972). Starting from an initial marginal $p_0(\hat{a})$, one alternates between updating the conditional channel,

$$p_{k+1}(\hat{a} \mid H) = \frac{p_k(\hat{a}) \exp[-\beta d_Z(H, \hat{a})]}{\sum_{a' \in \mathcal{A}} p_k(a') \exp[-\beta d_Z(H, a')]}, \tag{23}$$

and updating the marginal,

$$p_{k+1}(\hat{a}) = \sum_H P(H) p_{k+1}(\hat{a} \mid H). \tag{24}$$

Iterating equations (23) and (24) gives the fixed point for a chosen $\beta$. In the present framework, this means the minimum-information action channel is not merely characterized abstractly. It is computable whenever $P(H)$, $U(a \mid H, Z)$, and $\mathcal{A}$ are specified.

**Interpretation**

Equation (20) shows that the probability of selecting action $\hat{a}$ given support state $H$ is exponentially suppressed by the regret that action would incur. As $\beta$ increases, regret is penalized more strongly, and the channel approaches exact action sufficiency. As $\beta$ decreases, the action becomes less dependent on $H$, lowering information rate at the cost of higher regret.

The low-β limit should not be described as necessarily uniform. It is better described as becoming independent of $H$. In the zero-rate limit, the best default action is any

$$a_0 \in \arg\max_{a \in \mathcal{A}} \mathbb{E}_{\mathrm{H}}[U(\, a \mid H, Z\,)]\,. \tag{25}$$

Thus the zero-rate channel may concentrate on the marginal-utility-maximizing default action rather than spread uniformly across actions.

## 10. Exact limit of the rate-regret formulation

When $\varepsilon = 0$ and optimal actions are unique, any zero-regret action channel must select $\hat{A} = \pi^*(H, Z)$ almost surely.

Therefore the minimum exact rate is

$$R_Z(0) = I\big(H; \pi^*(H, Z)\big) = \mathsf{H}\big(\pi^*(H, Z)\big). \tag{26}$$

Thus the zero-regret rate is the entropy $\mathsf{H}$ of the optimal-action label induced by the consequence geometry. Equation (26) recovers Proposition 1 in information-theoretic form. Exact action sufficiency does not require preserving all information in $H$. It requires preserving the policy-equivalence class label.

## 11. Action margins and boundary geometry

The rate-regret formulation clarifies why consequence geometry matters. The amount of support information required depends on how probability mass in $\mathcal{H}$ lies relative to action boundaries.

**Definition 11. Action margin**

Define the action margin at $H$ as

$$m_Z(H) = U(\,\pi^*(H, Z) \mid H, Z\,) - \max_{a \neq \pi^*(H,Z)} \; U(\,a \mid H, Z\,). \tag{27}$$

Large margin means the optimal action is much better than the alternatives. Small margin means the support state lies near an action boundary.

**Proposition 5. A Markov bound for high-margin action errors**

For any stochastic action channel satisfying $\mathbb{E}\big[d_Z(H, \hat{A})\big] \leq \varepsilon$, and any $\gamma > 0$,

$$P\big(\hat{A} \neq \pi^*(H, Z) \text{ and } m_Z(H) \geq \gamma\big) \leq \frac{\varepsilon}{\gamma}. \tag{28}$$

**Proof**

If $\hat{A} \neq \pi^*(H, Z)$, then by definition $d_Z(H, \hat{A}) \geq m_Z(H)$.

On the event where $m_Z(H) \geq \gamma$, every nonoptimal action incurs regret at least $\gamma$. Therefore,

$$\mathbb{E}\left[d_Z(H, \hat{A})\right] \geq \gamma \cdot P\left(\hat{A} \neq \pi^*(H, Z),\ m_Z(H) \geq \gamma\right).$$

Since expected regret is at most $\varepsilon$, the result follows. ■

**Interpretation**

The bound in equation (28) shows that low-regret compression must preserve action labels reliably in high-margin regions. This is a feasibility bound, not an optimality claim. It applies to any stochastic action channel satisfying the expected-regret constraint, not only to the channel minimizing equation (19). Errors are most tolerable near low-margin boundaries, where alternative actions have similar value. Therefore the rate-regret curve depends on how probability mass in $\mathcal{H}$ is distributed relative to action boundaries, and on the regret incurred when those boundaries are crossed.

This is where uncertainty type enters the formal picture. Conflict, provenance fragility, and drift are not merely descriptive labels. They are dimensions along which consequence geometry can create action boundaries at fixed scalar confidence. If high-probability support states lie near boundaries induced by those dimensions, then scalar confidence may fail to preserve the distinctions needed for low-regret action.

Thus, for fixed ε, a consequence geometry that places substantial probability mass near action boundaries along conflict, provenance, or drift dimensions will generally require a higher minimum rate $R_Z(\varepsilon)$ when crossing those boundaries produces non-negligible regret. By contrast, boundaries that are sparse, low-probability, low-margin, or aligned with scalar confidence impose weaker rate demands.

## 12. Uniform sufficiency across consequence families

The fixed-$Z$ case is the cleanest result. But robust arbitration often requires performance across a family of consequence geometries.

Let $\mathcal{Z}$ be a family of consequence geometries, such as routine, threat, opportunity, and drift-sensitive regimes.

**Definition 12. Uniform exact action sufficiency**

A compression $g$ is uniformly exactly action-sufficient for $\mathcal{Z}$ if

$$\pi_g(g(H), Z) = \pi^*(H, Z) \quad \text{for all } H \in \mathcal{H} \quad \text{and all } Z \in \mathcal{Z}. \tag{29}$$

**Definition 13. Uniform $\varepsilon$-action sufficiency**

For a distribution $P(H)$, a compression $g$ is uniformly ε-action-sufficient for $\mathcal{Z}$ if

$$\mathbb{E}_{\mathrm{H}}\left[R_g(H, Z)\right] \leq \varepsilon \quad \text{for all } Z \in \mathcal{Z}. \tag{30}$$

Equations (29) and (30) distinguish exact uniform sufficiency from approximate uniform sufficiency across a family of consequence geometries. Uniform exact sufficiency is the special case in which compression preserves the full-information action for every support state and every consequence geometry. Uniform $\varepsilon$-sufficiency relaxes this requirement by allowing bounded expected regret in each consequence geometry.

**Corollary 2. Uniform scalar-confidence insufficiency**

If there exists any $Z \in \mathcal{Z}$ and any pair $H_1, H_2 \in \mathcal{H}$ such that $c(H_1) = c(H_2)$ but $\pi^*(H_1, Z) \neq \pi^*(H_2, Z)$, then scalar-confidence-only compression is not uniformly exactly action-sufficient for $\mathcal{Z}$.

**Proof**

Since $c(H_1) = c(H_2)$, scalar-confidence-only compression maps $H_1$ and $H_2$ to the same compressed state. The compressed policy must therefore choose the same action for both states under $Z$. But the full-information optimal actions differ. Thus scalar-confidence-only compression fails exact action sufficiency for that $Z$. Since uniform exact sufficiency requires exact sufficiency for every $Z \in \mathcal{Z}$, scalar-confidence-only compression is not uniformly exactly action-sufficient for $\mathcal{Z}$. ■

**Corollary 3. Uniform approximate scalar-confidence insufficiency**

Let $g_c(H) = c(H)$ be scalar-confidence-only compression. If there exists $Z \in \mathcal{Z}$ such that

$$\mathbb{E}_{\mathrm{H}}\left[R_{g_c}(H, Z)\right] > \varepsilon, \tag{31}$$

then, by equation (30), scalar-confidence-only compression is not uniformly $\varepsilon$-action-sufficient for $\mathcal{Z}$.

Equivalently, if scalar-confidence compression induces at least one compression cell whose expected regret under some $Z \in \mathcal{Z}$ pushes total regret above $\varepsilon$, then it fails uniform approximate sufficiency.

**Interpretation**

Confidence may be adequate in some environments. But once the family of consequence geometries contains regimes where conflict, provenance, drift, or other support dimensions alter the optimal action at fixed confidence, confidence alone cannot serve as a uniformly exact sufficient statistic. For approximate sufficiency, the question is not merely whether scalar confidence crosses an action boundary, but whether the regret induced by those crossings exceeds the tolerated loss $\varepsilon$. Thus scalar confidence can be approximately adequate in benign regimes while failing in threat, opportunity, or drift-sensitive regimes where the same collapsed support distinctions carry higher consequence.

## 13. Recurrent boundary of the action-channel reduction

The action-channel reduction is valid for the single-cycle problem because immediate regret depends only on the selected action. In recurrent arbitration, support may do more than select the immediate action. It may also update memory, recalibrate trust, alter future routing, or shape later support-resolution control.

In the recurrent architecture developed in Walsh (2026e), compressed support codes enter a memory-update loop. Let $M_t$ denote the arbitration memory at time $t$, understood as the accumulated record of prior support, actions, outcomes, and calibration history that can guide later arbitration. The basic point is that $M_{t+1}$ depends not only on the action selected, but also on the support structure preserved during arbitration and the outcome that follows. Later resolution control can then depend on that updated memory. As a result, information in $S_t$ that does not change the immediate action may still matter if it changes $M_{t+1}$ and thereby changes future arbitration.

The present paper does not solve that recurrent problem. It establishes the single-cycle base case. When only immediate action matters, support compression reduces to an action-channel problem. When support also affects memory updating and future resolution control, the relevant distortion is no longer one-step regret but long-horizon value loss. This is the natural target for a recurrent extension.

This caveat is important for the developmental setting. In the developmental simulations, low $\rho_t$ preserved coarse strength, medium $\rho_t$ preserved uncertainty type, and high $\rho_t$ preserved finer descriptors such as conflict, degradation, provenance, and novelty. The point was not merely immediate action selection, but how controllers learn to regulate support over time under changing consequence conditions (Walsh 2026c).

## 14. Discussion

### 14.1 Relation to rate-distortion theory

The rate-regret result uses standard rate-distortion machinery, but the distortion term has a different interpretation (Shannon 1959; Berger 1971). Classical rate-distortion theory asks how much information about a source must be preserved to reconstruct it within an acceptable fidelity criterion. In the present framework, the system does not need to reconstruct the full support state $H$. It needs to select an action with bounded regret. The distortion term is therefore not representational mismatch between $H$ and $S$. It is consequence-sensitive policy loss induced by selecting an action from compressed support.

This distinction matters because a great deal of support structure may be irrelevant to the immediate policy boundary. The exact result shows that zero-regret compression only needs to preserve the policy-equivalence class. The approximate result shows that bounded systems can permit some action-boundary errors when their expected cost is below tolerance. In both cases, the compression criterion is action sufficiency rather than reconstruction fidelity.

### 14.2 Relation to the information bottleneck

The structure also resembles the information bottleneck (Tishby et al. 2000). The information bottleneck compresses a source while preserving information relevant to a target variable. Support sufficiency similarly asks which aspects of $H$ must be preserved for downstream use. The difference is the relevance criterion. In the standard information bottleneck, relevance is usually predictive information about a target variable. In support sufficiency, relevance is policy adequacy under consequence geometry.

Thus, the sufficient statistic is not whatever best predicts an external label in general. It is whatever preserves the action-equivalence partition, or its bounded-regret approximation, under $Z$. This shifts the bottleneck from predictive sufficiency to policy sufficiency. The same support state may be compressible in one consequence geometry and not in another, because the action boundaries induced by $Z$ determine which distinctions matter.

### 14.3 Relation to rational inattention and discrete choice

The framework is also adjacent to rational inattention and information-constrained discrete choice. Rational inattention studies how agents allocate limited information-processing capacity before choosing actions, and discrete-choice versions show how information-cost optimization can yield logit or Gibbs-like stochastic choice rules. Matějka and McKay derive a rational-inattention foundation for multinomial logit choice, while Caplin and Dean characterize revealed-preference conditions for costly information acquisition (Matějka and McKay 2015; Caplin and Dean 2015). These models are mathematically closer to the present rate-regret formulation than generic rational-inattention references alone suggest.

The distinction is therefore not that the present paper introduces the first information-constrained action model. It does not. The distinction is the object being compressed and the interpretation of the distortion term. Rational-inattention models typically ask how an agent should allocate information about states, options, or payoffs under information costs. The present framework asks what distinctions in an already formed support state must survive an arbitration bottleneck. The source is not simply an external state or option value. It is a full support state $H$ containing evidential strength, conflict, provenance, drift, and memory-conditioned context. The distortion is not generic loss from bounded choice, but regret relative to the full-information policy under a consequence geometry.

Thus, action-sufficient compression should be viewed as complementary to rational inattention rather than as a replacement for it. Rational inattention primarily concerns what information to acquire, attend to, or process before choice. Support sufficiency concerns what support structure to preserve once evidence has already shaped $H$. These questions can interact, but the formal target here is preservation of action-relevant support distinctions at the point of arbitration.

**14.4 Relation to information-theoretic bounded rationality**

The rate-regret formulation also overlaps with information-theoretic bounded rationality and information-theoretic accounts of decision and action. Tishby and Polani frame decision and action within the perception-action cycle using information-theoretic quantities (Tishby and Polani 2011). Ortega and Braun formulate bounded decision-making as a tradeoff between expected utility and information-processing costs (Ortega and Braun 2013). Genewein and colleagues extend this line to abstraction and hierarchical decision-making under information-processing constraints (Genewein et al. 2015).

The present framework should be located inside that broader family. Its contribution is not the general idea that bounded agents trade utility against information. Its contribution is to specify a support-sufficiency instance of that tradeoff. The source variable is the support state $H$, not merely an environmental state. The relevant compression loss is policy regret induced by collapsing support distinctions that consequence geometry makes action-relevant. This allows conflict, provenance fragility, and drift to be treated as information-bearing dimensions whose value depends on how they alter action boundaries.

**14.5 Open problem: recurrent support sufficiency**

The rate-regret formulation developed above is single-cycle. In recurrent arbitration, support can matter even when it does not change the immediate action, because it may update memory, recalibrate trust, alter future routing, or shape later support-resolution control. The open problem is therefore to generalize the rate-regret characterization from immediate policy regret to long-horizon value loss under memory dynamics. A closely related formal precedent is Tiomkin and Tishby's information-theoretic Bellman treatment of causal information and value in Markov decision processes, which similarly links information constraints to recurrent value structure (Tiomkin and Tishby 2017).

A natural future objective would condition the rate term on arbitration memory, since $M_t$ already carries information from prior support, action, and outcome histories. One candidate finite-horizon formulation is

$$\min_{\{p(S_t|H_t, M_t)\}_{t=0}^{T}} \mathbb{E}\left[\sum_{t=0}^{T} \delta^t I(H_t; S_t \mid M_t)\right] \text{ subject to } \mathbb{E}\left[V^*(M_0, H_0, Z_0) - V_g(M_0, H_0, Z_0)\right] \leq \varepsilon. \quad (32)$$

In equation (32), the distortion term is no longer one-step regret but long-horizon value loss under memory-update dynamics. This recurrent problem is not a minor corollary of the single-cycle proposition. Once $S_t$ affects $M_{t+1}$, support distinctions can matter even when they do not change the immediate action.

This would formally connect the single-cycle proposition to the developmental support-resolution setting, where $S_t$ can matter because it shapes how $M_t$, future $\rho_t$, and later policy selection evolve. In the degenerate case where memory is absent or irrelevant and only immediate action matters, the recurrent objective should reduce to the single-cycle rate-regret formulation developed here.

### 14.6 Summary and conclusion

The single-cycle result can now be stated compactly. Support sufficiency is action-sufficient compression. For fixed consequence geometry, the exact target is the quotient of support space by policy equivalence. For bounded systems, the approximate target is the minimum-information action channel whose expected policy regret remains within tolerance. Content and scalar confidence are sufficient only when their induced partitions refine, or approximate closely enough, the action-equivalence partition. When they collapse support states that require different actions, they are not action-sufficient.

This paper formalized single-cycle support sufficiency as a consequence-sensitive compression problem. The source is the full support state $H$, the reproduction alphabet is the action set $\mathcal{A}$, and the distortion measure is policy regret under $Z$. The resulting rate-regret formulation applies standard rate-distortion machinery to a specific decision-theoretic object: the support distinctions that consequence geometry makes action-relevant.

The proposition sequence clarifies the normative basis of support-sensitive arbitration. Robust decision-making does not require preserving all support, but it does require preserving the distinctions that keep the system on the correct side of action boundaries. This explains why selected content and scalar confidence can be adequate in benign regimes yet fail in threat, opportunity, or drift-sensitive regimes. Their adequacy depends on whether their induced partitions preserve the policy-relevant geometry of the support space.

The recurrent extension remains a boundary condition rather than a solved corollary. In dynamic systems, support may matter not only for immediate action, but also for memory updating, trust calibration, future routing, and later support-resolution control. The present paper supplies the single-cycle case that such a recurrent theory should recover. When memory is degenerate and only immediate action matters, support sufficiency reduces to action-sufficient rate-regret compression.

## Declarations

### Funding

This research did not receive any specific grant from funding agencies in the public, commercial, or not-for-profit sectors.

### Competing interests

The author declares that he has no known competing financial interests or personal relationships that could have appeared to influence the work reported in this paper.

### Data availability

No empirical data were generated or analyzed for this theoretical article.

### Author contributions

Mark Walsh: Conceptualization, formal analysis, methodology, writing, original draft, writing, review and editing.

### Declaration of generative AI and AI-assisted technologies in the manuscript preparation process

During the preparation of this work, the author used OpenAI's ChatGPT to support wording revision and formatting review. After using this tool, the author reviewed, edited, and verified the content as needed and takes full responsibility for the content of the submitted manuscript.